\documentstyle[aps,amsfonts]{revtex}
\def\vec#1{\bbox{#1}}
\def\ctp{{\mbox{\footnotesize CTP}}}
\def\tfd{{\mbox{\footnotesize TFD}}}
\def\Av#1{\left\langle #1 \vphantom{\int}\right\rangle}
\def\TO#1{\mbox{\large T}\left[ #1 \right]}
\def\TOt#1{\widetilde{\mbox{\large T}}\left[ #1 \right]}
\def\i{{\mathrm i}}
\begin{document}
\title{Comment on the diagonalization of Green functions
       \thanks{Work supported by GSI, Preprint No. 95-12 (1995)}}
\author{P.A.Henning\thanks{
  Electronic mail: P.Henning\@ gsi.de,
  http://www.gsi.de/$\widetilde{ }$phenning/henning.html}}
\address{Theoretical Physics,
        Gesellschaft f\"ur Schwerionenforschung GSI\\
        P.O.Box 110552, D-64220 Darmstadt, Germany}
\maketitle
\begin{abstract}
Real-time thermal field theory is known in the two flavors
``closed-time path formalism'' and ``thermo field dynamics''.
With a certain choice of parameters the full two-point
functions of these models are identical, hence a scheme
to diagonalize the Green function in one model
can be applied to the other as well. This allows
to compare the diagonalization schemes that have been
discussed in the recent literature in order to select
the simplest one.  Such comparison speaks in favor
of a diagonalization method which
is motivated from thermo field dynamics.
\end{abstract}
In real-time quantum field theory for statistical systems, i.e., for
thermal as well as for non-equilibrium states, the two-point
Green functions acquire a $2\times 2$ matrix structure. Two flavors exist
for such a theory: The Schwinger-Keldysh closed-time path formalism
(CTP, \cite{SKF}) and thermo field dynamics (TFD, \cite{TFD}).

These matrix valued Green functions (propagators) contain spurious
information that unnecessarily complicates their use in a perturbative
expansion. This unneeded information may be removed by matrix diagonalization.
The fact that different diagonalization methods are possible has led
to a dispute in the recent literature \cite{X95,hu92,AB92,EW92};
the present work is aimed at a clarification of this question,
thereby uniting the different approaches.

To this end, we concentrate on an interacting scalar quantum field
in a thermal equilibrium state and note that similar
considerations apply for fermions.
The first question to be settled is, whether the Green functions under
dispute are different. By definition, the propagator matrix
in the CTP formalism is \cite{LW87}
\begin{equation}\label{ctpp}
D^{(ab)}_\ctp(x,x^\prime)  = -\i\left( {
   \array{lr}
    \Av{ \TO{\phi_x \phi_{x^\prime}} } &
    \Av{ \phi_{x^\prime}\phi_x  \vphantom{\int}      } \\[2mm]
    \Av{ \phi_x\phi_{x^\prime} \vphantom{\int}      } &
    \Av{ \TOt{\phi_x\phi_{x^\prime}} }
   \endarray} \right)
\;,\end{equation}
where $\Av{\cdot}$ denotes the statistical average, $\TO{\cdot}$
the time ordered and $\TOt{\cdot}$ the
''anti-time ordered'' product, i.e.,
\begin{eqnarray}\label{atpb}\nonumber
\TO{\phi_x \phi_{x^\prime}}  &= &
\Theta(x^0-x^{0\prime})\,\phi_x \phi_{x^\prime} +
\Theta(x^{0\prime}-x^0)\,\phi_{x^\prime} \phi_x\\
\TOt{\phi_x \phi_{x^\prime}}  &= &
\Theta(x^{0\prime}-x^0)\,\phi_x \phi_{x^\prime} +
\Theta(x^0-x^{0\prime})\,\phi_{x^\prime} \phi_x
\;.\end{eqnarray}
In the TFD formalism, the canonical commutation relations
have two different commuting representations $\phi_x$
and $\widetilde{\phi}_x$, the matrix valued propagator is
\begin{equation}\label{tfdp}
D^{(ab)}_\tfd(x,x^\prime)
   =-\i\left( {
   \array{ll}
    \Av{ \TO{\vphantom{\int}\phi_x \phi_{x^\prime}} }&
    \Av{ \TO{\vphantom{\int}\phi_x\widetilde{\phi}_{x^\prime}} }\\[2mm]
    \Av{ \TO{\vphantom{\int}
  \widetilde{\phi}_x \phi_{x^\prime}} }&
    \Av{ \TO{\vphantom{\int}
  \widetilde{\phi}_x\widetilde{\phi}_{x^\prime}} }
   \endarray} \right)
\;.\end{equation}
TFD is conceptually different from CTP, in the sense that it contains
a Bogoliubov transformation of quantum fields at the operator level.
This transformation contains three parameters, for the purpose
of the present paper only one aspect is interesting:
{\em For a certain choice of TFD parameters, the two matrices above
are identical~}\cite{hu92}. By adopting this special
choice ($\alpha=1$), we will henceforth omit a distinction
between $D_\ctp$ and $D_\tfd$, and conclude that they may be diagonalized
by the same matrix transformation.

We now turn to the task of isolating the spurious information
hidden in these propagators. For convenience, we
perform a Fourier transformation to the momentum variable $p=(p_0,\vec{p})$,
with respect to the difference of the space-time coordinates.

The goal is to formulate a perturbative expansion
in terms of the retarded and advanced functions
\begin{eqnarray}\label{dar}\nonumber
D^{R}(p) &=& D^{11}(p)-D^{12}(p)\\
D^{A}(p) &=& D^{11}(p)-D^{21}(p)
\;=\;\left( D^R(p) \right)^\star
\;,\end{eqnarray}
since they are free of poles on the physical complex energy sheet other
than on the real axis. By inspection of (\ref{ctpp}) and ({\ref{tfdp})
one may easily realize that
\begin{eqnarray}
\nonumber
D^{11}(p) + D^{22}(p) & = \;D^{12}(p) + D^{21}(p)\\
\nonumber
D^{11}(p) & = \; -\left( D^{22}(p) \right)^\star&\\
D^{12}(p) & = \; -\left( D^{12}(p) \right)^\star&
\;,\;\;\;\;\;\;\;
\label{line}
D^{21}(p) \; = \; -\left( D^{21}(p) \right)^\star
\;,\end{eqnarray}
where $\star$ denotes complex conjugation. Apart from these
trivial relations we also know, that in a thermal equilibrium state
the Kubo-Martin-Schwinger boundary condition holds,
\begin{equation}\label{KMS}
(1 + n_B(p_0)) D^{12}(p) - n_B(p_0) D^{21}(p) =0
\;,\end{equation}
with the Bose-Einstein function
\begin{equation}
n_B(E) = \frac{1}{\displaystyle \exp(\beta (E-\mu)) -1}
\;\end{equation}
at inverse temperature $\beta$ and chemical potential $\mu$
($\mu=0$ for the real scalar field).

{}From these relations follows that there exists a
nonsingular $2\times2$ matrix $V(p_0)$ such that
\begin{equation}\label{diactp}
V(p_0)\, D(p)\, V(p_0) =
\left( { \array{lr}  D^{F}(p) & 0\\
                     0        & -\left(D^{F}(p)\right)^\star
         \endarray} \right)
\;,\end{equation}
where
\begin{equation}\label{df}
D^{F}(p)=\Theta(p_0)\,D^{R}(p)
                  +\Theta(-p_0)\,D^{A}(p)
\;.\end{equation}
(this ``Feynman''-like propagator is not causal
in a state with nonzero $n_B(p_0)$).

This matrix
transformation {\em diagonalizes\/} the full boson propagator.
The transformation matrix is well known for some time
(see e.g. \cite[eq. (2.4.31)]{LW87}), here we adopt the notation
of ref. \cite{X95}:
\begin{equation}\label{ndef}
V(p_0) = \left( {
\array{lr} \cosh(\theta) & -\exp[\beta(p_0-\mu)/2]\sinh(\theta)) \\
          -\exp[-\beta(p_0-\mu)/2]\sinh(\theta) & \cosh(\theta)
 \endarray} \right)
\;,\end{equation}
with hyperbolic functions
\begin{eqnarray}\nonumber
\cosh(\theta) & = & \sqrt{ \Theta(p_0) [ 1+ n_B(p_0) ]
                          -\Theta(-p_0) n_B(p_0)}\\
\sinh(\theta) & = & \sqrt{ \Theta(p_0) n_B(p_0)
                          -\Theta(-p_0)[1+ n_B(p_0)]}
\;.\end{eqnarray}
Thus, after diagonalization the only temperature dependence is
located in the matrices $V$ -- and we may absorb
them into the interaction vertices of a theory, which are then linked
by retarded or advanced propagators. The goal of isolating the spurious
information hidden in the propagator from its field theoretical
content has been reached. However, three questions are unanswered yet:
\begin{enumerate}
\item Does the matrix (\ref{ndef}) represent the simplest possible choice ?
\item
What is the physical meaning of this diagonalization procedure ?
\item How does one handle the
diagonalization task in non-equilibrium states ?
\end{enumerate}
To answer the first question, we reconsider the relations (\ref{line})
and (\ref{KMS}). Apart from the diagonalization procedure outlined above, they
also guarantee a different way to reach our goal: There exists at least one
nonsingular $2\times2$ matrix ${\cal B}$ such that
\begin{equation}\label{diatfd}
{\cal B}[n_B(p_0)]\; D(p)\; \tau_3\; \left({\cal B}[n_B(p_0)]
\vphantom{\int}\right)^{-1} =
\left( { \array{lr}  D^{A}(p) & 0\\
                     0                   & D^{R}(p)
         \endarray} \right)
\;,\end{equation}
where $\tau_3=\mbox{diag}(1,-1)$. Hence, also this matrix
transformation {\em diagonalizes\/} the full boson propagator
\cite{hu92,UY92,habil}.

In fact, more than one such matrix ${\cal B}$ exists, but
one may chose a particularly simple form linear in
the Bose-Einstein function $n(p_0)$
\begin{equation}\label{bdef}
{\cal B}[n] = \left( {
\array{ll} \left(1 + n\right) & n \\
           -1           & 1 \endarray} \right)
\;.\end{equation}
{}From the continuum of possible diagonalization methods of the full
propagator, the above choice clearly is one of the simplest --
whereas the matrix $V$ in (\ref{ndef}) is a highly nonlinear
function of $n(p_0)$.

To answer the second question, we consider a simple harmonic oscillator with
hamiltonian $H = \omega\, a^\dagger\, a$,
immersed in a heat bath of inverse temperature $\beta$.
It is well known, that one may describe this oscillator equally well in
terms of particle states or hole states, i.e., the
Liouville operator governing the time evolution of the
statistical system has a symplectic symmetry.
Without elaboration at this point we note
that its symmetry group is the two-dimensional symplectic group Sp(2)
(see \cite{H93,habil} for a more complete discussion).

In a thermal equilibrium state, we also know the statistical
operator to be $W = \exp(-\beta H)$. It follows that
\begin{eqnarray}\label{com}\nonumber
(1 + n_B(\omega))\, a\, W  & - n_B(\omega)\, W\, a &=0\\
(1 + n_B(\omega))\, W\, a^\dagger  & -  n_B(\omega)\, a^\dagger\, W &=0
\;,\end{eqnarray}
which is nothing but the Kubo-Martin-Schwinger
condition for this oscillator.
Thus, although the creation and annihilation operators of this simple system
do not commute with the statistical operator, we find that
a certain {\em linear combination\/} of right-acting and left-acting
operators (well defined in Liouville space \cite{H93})
annihilates the density matrix.

The corresponding orthogonal linear combination
creates an excitation of the system. It is a linear
combination of particle and hole state, which propagates through
the system without ``feeling'' the thermal background.
Thus, by introducing these
``thermal quasi-particles'' the statistical information has
been separated from the time evolution problem.

The relation between ordinary creation/annihilation operators
and these unusual linear combinations is mediated by the
matrices ${\cal B}[n(\omega)]$ from eq. (\ref{bdef}).
Hence we may identify the diagonalization according
to eq. (\ref{diatfd}) {\em physically\/} as the transformation to
a linear combination of particle and hole state that have
retarded/advanced boundary conditions in time.

It remains to answer the third question, i.e., what one does
for non-equilibrium states. In this case, the Fourier
transform with respect to coordinate differences
leaves us with an additional dependence on $X=(x+x^\prime)/2$.
In this mixed representation, the
relations (\ref{line}) between the
matrix elements of the two-point functions still prevail,
but the KMS boundary condition (\ref{KMS}) is {\em not\/} valid.

We label the non-equilibrium propagators by a $\widetilde{\;}$-sign,
i.e. $\widetilde{D}^{ab}\equiv\widetilde{D}^{ab}(X,p)$,
and use the easily established fact that for {\em arbitrary}
parameter $N\equiv N(X,p)$ the matrix
$\widetilde{{\cal B}}[N]$ according to (\ref{bdef}) leads to
\begin{equation}\label{dianeq}
\widetilde{{\cal B}}[N]\, \widetilde{D}(X,p)\, \tau_3\,
\left(\widetilde{{\cal B}}[N]\right)^{-1} =
\left( { \array{lr}  \widetilde{D}^{A} &\;\;
  \left(N \widetilde{D}^{12}\vphantom{\int}
-(1+ N)\widetilde{D}^{21}\right) \\
                     0                   & \widetilde{D}^{R}
         \endarray} \right)
\;.\end{equation}
Hence, this procedure transforms the full Green function to triangular form.
One might then ask, whether there exists a special choice
for $N\equiv N(X,p)$ such that the off-diagonal element of this matrix
also vanishes, i.e. such that
\begin{equation}\label{tpe}
  N(X,p) \widetilde{D}^{12}(X,p)
-\left(1+ N(X,p)\right)\widetilde{D}^{21}(X,p) = 0
\;.\end{equation}
This can be answered by inserting the propagator into the
full Schwinger-Dyson equation, the result is
a differential equation for $N(X,p)$. On close
inspection it is identified with  a {\em transport equation}, i.e.,
a close relative of the Boltzmann equation for the
quantity $N(X,p)$ \cite{habil,h94gl3}.

Naturally, transport equations are also obtainable using a different
diagonalization scheme. However, the differential equations
then are {\em much\/} more complicated and thus
their relation to the Boltzmann equation or any other
{\em known\/} transport equation is not easily established.

To conclude the present paper: As already stated in ref. \cite{hu92},
a diagonalization of two-point functions is easily possible
in the closed-time path formalism as well as in thermo field dynamics --
which is a trivial fact since the propagators can be made identical.

As pointed out in ref. \cite{X95} as well as in \cite{hu92},
the diagonalization matrices that occur {\em naturally} are
$V$ from (\ref{ndef}) in CTP, but ${\cal B}$ from (\ref{bdef}) in TFD.
For simplicity one should therefore make the obvious choice
of the diagonalization scheme according to eq. (\ref{diatfd})
{\em also in the CTP formalism\/}. Especially for
non-equilibrium states such simplicity is {\em required\/}
to obtain a meaningful interpretation of the diagonalization
condition -- independently of the reader's preference for
one of the two flavors of real-time statistical quantum field theory.

The situation described here bears a close analogy to the
fixing of a gauge: In equilibrium states,
the choice of diagonalization matrices does not influence physical
quantities -- but the {\em calculational effort\/} greatly
depends on this choice. It has been shown,
that this is indeed more than an analogy, i.e.,
that choosing a special form of the transformation matrices
indeed corresponds to a gauge fixing \cite{habil}.


\begin{thebibliography}{99}
\bibitem{SKF}{
    J.Schwinger,
    J.Math.Phys. {\bf 2} (1961) 407;\\[1mm]
    L.V.Keldysh, Zh.Exsp.Teor.Fiz. {\bf 47} (1964) 1515 and
    JETP {\bf 20} (1965) 1018}
\bibitem{TFD}{
    T.Arimitsu and H.Umezawa,
    Prog.Theor.Phys. {\bf 77} (1987) 32 and 53;\\[1mm]
    H.Umezawa,
    {\em Advanced Field Theory: Micro, Macro and Thermal Physics}\\
    (American Institute of Physics, 1993)}
\bibitem{X95}{
    H.Xu, Phys.Lett.{\bf B342} (1995) 219}
\bibitem{hu92}{
    P.A.Henning and H.Umezawa,\\
    Nucl.Phys. {\bf B417} (1994) 463 and
    Phys.Lett. {\bf B303} (1993) 209}
\bibitem{AB92}{
    P.Aurenche and T.Becherrawy,
    Nucl.Phys. {\bf B379} (1992) 259}
\bibitem{EW92}{
    M.A.van Eijck and Ch.G. van Weert,
    Phys.Lett. {\bf B278} (1992) 305}
\bibitem{LW87}{
    N.P.Landsman and Ch.G.van Weert,
    Phys.Rep. {\bf 145} (1987) 141}
\bibitem{UY92}{
    H.Umezawa and Y.Yamanaka, Mod.Phys.Lett. {\bf A7} (1992) 305}
\bibitem{habil}{
    P.A.Henning,\\
    {\em TFD for quantum fields with continuous mass spectrum},\\
    Physics Report (to appear in February 1995)}
\bibitem{H93}{
    M.Hirokawa, Ann.Phys. {\bf 223} (1993) 1}
\bibitem{h94gl3}{
    P.A.Henning,
    Nucl.Phys. {\bf A582} (1995) 633}
\end{thebibliography}
\end{document}